# Semimetal-semiconductor transition and giant linear magnetoresistances in three-dimensional Dirac semimetal $Bi_{0.96}Sb_{0.04}$ single crystals

Z. J. Yue,[1] X. L. Wang[1a)] and S. S. Yan[2]

[1]*Spintronics and Electronic Materials Group, Institute for Superconducting and Electronic Materials, Australian Institute for Innovative Materials, University of Wollongong, Squires Way, North Wollongong, NSW 2500, Australia*

[2]*School of Physics, State Key Laboratory of Crystal Materials, Shandong University, Jinan 250100, People's Republic of China*

Three-dimensional (3D) Dirac semimetals are new quantum materials and can be viewed as 3D analogues of graphene. Many fascinating electronic properties have been proposed and realized in 3D Dirac semimetals, which demonstrates their potential applications in next generation quantum devices. Bismuth-antimony $Bi_{1-x}Sb_x$ can be tuned from a topological insulator to a band insulator through a quantum critical point at $x \approx 4\%$, where 3D Dirac fermions appear. Here, we report on a magnetotransport study of $Bi_{1-x}Sb_x$ at such a quantum critical point. An unusual magnetic-field induced semimetal-semiconductor phase transition was observed in the $Bi_{0.96}Sb_{0.04}$ single crystals. In a magnetic field of 8 T, $Bi_{0.96}Sb_{0.04}$ single crystals show giant magnetoresistances of up to 6000% at low-temperature, 5 K, and 300% at room-temperature, 300 K. The observed magnetoresistances keep linear down to approximate zero-field when the temperature is below 200 K. Our experimental results are not only interesting for the fundamental physics of 3D Dirac semimetals, but also for potential applications of 3D Dirac semimetals in magnetoelectronic devices.

Gapless Dirac materials have fascinating electronic properties and numerous potential practical applications in future spintronic, electronic, and optical devices.[1] Graphene is the most well-known Dirac material and demonstrates extraordinarily high mobility, thermal conductivity, and mechanical strength.[2] These fascinating properties in graphene have triggered numerous fundamental and technological studies.[3, 4] 3D Dirac semimetals can be viewed as the 3D generalization of graphene, and they represent a new state of quantum matter.[5] In these systems, 3D Dirac points are protected by crystal symmetry and have linear energy dispersion in any momentum direction. Recently, 3D Dirac semimetals were theoretically proposed and experimentally realized in $Na_3Bi$ and $Cd_3As_2$ using first-principle calculation and angle resolved photoemission spectroscopy (ARPES), respectively.[6, 7] Bismuth-antimony $Bi_{1-x}Sb_x$ alloy is well-known as an excellent thermoelectric material and topological insulator.[8, 9] It can be tuned from a topological insulator to a band insulator through a critical point $x \approx 4\%$.[10, 11, 12] At such a critical point $x \approx 4\%$, $Bi_{1-x}Sb_x$ becomes a 3D Dirac semimetal phase.[13]

Many fascinating electronic transport properties have been predicted theoretically for 3D Dirac semimetals.[14] With tiny pockets in the Fermi surface, linear quantum magnetoresistances (MR) are expected in these systems at the extreme quantum limit.[15, 16] Recent transport experiments have demonstrated quantum magnetoresistance, quantum oscillations, and ultra-high mobility, as well as suppression of electron backscattering in $Bi_{1-x}Sb_x$ and $Cd_3As_2$.[12, 17, 18] Here, we report on a magnetic-field induced unusual semimetal-semiconductor phase transition at the critical point of $Bi_{1-x}Sb_x$. We observed giant low-temperature MR and large room-temperature MR in 3D Dirac semimetal phase of $Bi_{1-x}Sb_x$ single crystals. The observed low-temperature MR remains linear down to near zero-field and demonstrates possible potential for magnetoelectronic devices.

The $Bi_{1-x}Sb_x$ bulk single crystals were grown in a high temperature furnace using the same method stated in reference 9. The stoichiometric mixtures of high-purity Bi and Sb elements were sealed in a vacuum quartz tube to avoid oxidations. The mixtures were heated to 650 °C, and cooled to 270 °C over a period of five days. Then, they were annealed for seven days at 270 °C. X-ray diffraction measurements demonstrated that the samples were single phase, and have a rhombohedral crystal structure. The Bi$_{0.96}$Sb$_{0.04}$ samples used for the magnetotransport experiments were cleaved along the (001) plane from bulk crystals with 99.99% purity.

Four-probe transport measurements were performed on a rectangular sample with dimensions of 2 × 2 × 0.2 mm$^3$ between 5 and 350 K using a Quantum Design 9 T Physical Properties Measurement System (PPMS). The resistance was obtained by applying an electric current *I* (typically 1 mA) through the two outer contacts and monitoring the voltage drop *V* between the two inner contacts (typical spacing 1 mm). The current *I* was applied in the plane, and the magnetic field *B* was applied perpendicular to the current direction. And the *B* also was kept along the *c* axis and is perpendicular to the cleaved (001) plane. Similar measurement configurations can be found in our former works.[19, 20]

Figure 1(a) shows the temperature dependence of the measured resistance in Bi$_{0.96}$Sb$_{0.04}$ single crystals for a series of applied magnetic fields. In 0 T and 0.1 T, the resistivity, $\rho$, monotonically decreases with decreasing temperature, and the Bi$_{0.96}$Sb$_{0.04}$ single crystal behaves as a metal. On the contrary, in a higher field (*B* > 0.1 T), the resistivity increases monotonically with decreasing temperature, and the Bi$_{0.96}$Sb$_{0.04}$ single crystal behaves as a semiconductor. The temperature dependence of the MR was extracted and is displayed in Figure 1(b). The MR increases with decreasing temperature and reaches its maximum value of 6000% at 5 K and 8 T. The magnetic field induced phase transition from a semimetal to a semiconductor suggests a change in the band structure and a possible presence of a band gap. Based on the thermal activation energies for electrical conduction and the well-known equation of $\rho(T) = \rho_0 \exp(E_g/2k_BT)$, the thermal activation energy gap was extracted through drawing Ln($\rho$) as a function of $T^{-1}$, as shown in Figure 1(c).[21, 22] The induced thermal activation energy gap shows magnetic field dependence and reaches 117 meV in 8 T, as shown in Figure 1(d).

Such an unusual semimetal-semiconductor transition was observed in conventional semimetals, such as Bi and graphite.[23, 24] Recently, it has also been observed in transition-metal dichalcogenide WTe$_2$ and 3D weyl semimetals NbP.[25, 26] In contrast with WTe$_2$, 3D Dirac semimetal Bi$_{0.96}$Sb$_{0.04}$ demonstrates a transition at higher temperature lower magnetic fields. These quantum materials share the common features such as low carrier density, small effective mass, and an equal number of electrons and holes (compensation). As a result, similar to Bi and graphite, 3D Dirac semimetal Bi$_{0.96}$Sb$_{0.04}$ may also obey unique energy scale constraints that lead to observed semimetal-semiconductor transition accompanied by large linear MR.[27]

On the other hand, strong interaction could result in a breaking of dynamical chiral symmetry and a semimetal-semiconductor transition.[28] Electron-electron interaction driven metal-insulator phase transition has been proposed in a Dirac semimetal.[29] The interactions in 3D Dirac semimetals $Bi_{0.96}Sb_{0.04}$ may also play an important role in the semimetal-semiconductor transition when the system locates in high magnetic fields.[30] In addition, disorder scattering, electron localization, and field-induced change of band structure and electron and hole pockets moves may also play a role in the phase transition.[31] However, further quantitative theoretical investigations are needed for an accurate explanation of our observations.

Figure 2(a) shows the magnetic field dependence of the transverse MR in $Bi_{0.96}Sb_{0.04}$ single crystals. The maximum MR ratio exceeds 6000% at $T = 5$ K and $B = 8$ T. The high-field MR always displays linear field dependence. The MR reaches 300% at 300 K and 8 T, and shows no trend towards saturation. The well-established Kohler's rule suggests that the MR of a material is a universal function of $B$, as a result of the Lorentz force deflection of carriers. At high field, most materials show saturation of MR. Therefore, such a giant and linear MR in our $Bi_{1-x}Sb_x$ crystals violates Kohler's rule.

Figure 2(b) shows the low-field linear MR in macroscopically inhomogeneous $Bi_{0.96}Sb_{0.04}$ single crystals. The linear dependence on magnetic field emerged from near zero-field and continued up to room temperature. Quantum-MR theory was developed by Abrikosov to explain the observations of giant linear MR.[32] The model is based on the assumption that the systems are basically gapless semiconductors with a linear energy spectrum, and only one Landau Level (LL) participates in the conductivity.[32] $Bi_{0.96}Sb_{0.04}$ has small pockets in its Fermi surface with a small effective mass and shows metallic behavior in zero field.[10] The estimated carrier concentration and mobility from Hall measurements are $2 \times 10^{18}$ cm$^{-3}$ and $2.2 \times 10^5$ cm$^2$/Vs at 5 K, respectively. However, based on the reported magnetotransport studies, 3D Dirac semimetal $Bi_{0.97}Sb_{0.03}$ could not reach the ultra quantum limit of lowest LL, even when the magnetic field was increased up to 60 T.[12] With a magnetic field of 8 T, our $Bi_{0.96}Sb_{0.04}$ system also unlikely reaches the quantum limit. Therefore, $Bi_{0.96}Sb_{0.04}$ system is still within a semi-classical limit and there are multiple LLs participating in the quantum transport.

Large linear MR can also occur in semiconductors with parabolic band structures as a consequence of disorders.[21] Macroscopic disorder was used to explain the non-saturating MR in doped silver chalcogenides.[33] The low-field MR was found to be linear in macroscopically inhomogeneous InSb.[21] With gross inhomogeneities, distorted current paths could be generated in the semiconductors. The observed nonsaturating linear MR in 3D Dirac semimetal $Cd_3As_2$ was also attributed to disorder effects and mobility fluctuation.[34] Recently, a new semi-classical theory, named "guiding center linear MR", was proposed to explain the observed linear and non-saturating MR in 3D Dirac semimetals.[35] Such a guiding center linear MR occurs when disorder potentials are stronger than the Debye frequency. This kind of linear MR is predicted to survive up to room temperature and beyond. The disorders in $Bi_{0.96}Sb_{0.04}$ are not weak and the mobility is high, the observed linear MR in $Bi_{0.96}Sb_{0.04}$ system is attributable to the disorder-induced guiding center linear MR.

Large MR has been observed in topological insulators, including $Bi_2Te_3$ nanosheets and electrodeposited bismuth thin films.[36, 23] Giant linear MR has also been observed in the narrow-gap nonmagnetic semiconductors $Ag_{2\pm\delta}Te$ and $Ag_{2\pm\delta}Se$.[37, 38] In Figure 3(a) and (b), we compare the magnetic field and temperature dependences of the absolute values of MR in $Bi_{0.96}Sb_{0.04}$ with those in $Ag_{2+\delta}Se$, $Bi_2Te_3$, InSb, $Ag_{2+\delta}Te$, and $La_{0.75}Ca_{0.25}MnO_3$. Compared with $Bi_2Te_3$ nanosheets, $Bi_{0.96}Sb_{0.04}$ displays low-field linear MR. Compared with silver chalcogenides, $Bi_{0.96}Sb_{0.04}$ demonstrates larger low-field linear MR at room temperature.

In conclusion, we investigated the Dirac semimetal phase of $Bi_{1-x}Sb_x$ and observed a magnetic field induced semimetal-semiconductor transition and giant linear magnetoresistance. The magnetic field-dependent thermal activation energy gap reaches up to 117 meV at 8 T. Large linear magnetoresistances of up to 300% at room temperature and 6000% at 5 K were observed at 8 T. The low-field MR remains linear down to near 0 T and up to room-temperature. Our work may lead to further investigations of unusual semimetal to semiconductor transition and find its potential applications in magnetoelectronic devices.

**Acknowledgements**

This work is partially supported by the Australian Research Council under a Discovery project (ARC Discovery, DP130102956). This work is also supported by 111 projects B13029.

**Figures**

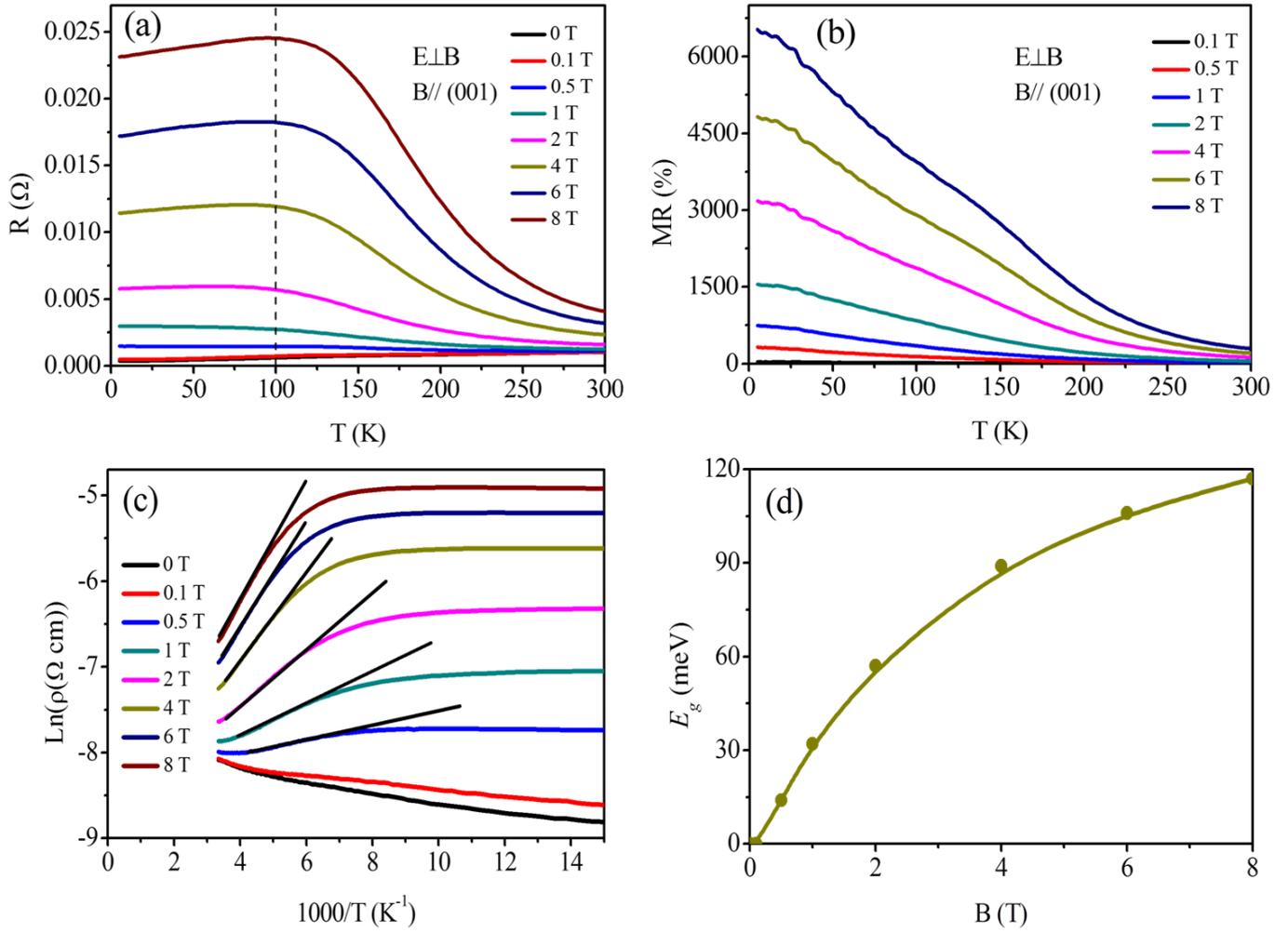

**Figure 1** (a) Resistance of Bi$_{0.96}$Sb$_{0.04}$ single crystals as a function of temperature in different magnetic fields ranging from 0 T to 8 T. (b) Temperature dependence of MR in Bi$_{0.96}$Sb$_{0.04}$ single crystals at different magnetic fields. (c) Ln($\rho$) as a function of $T^{-1}$. (d) The field induced thermal activation energy gap as a function of applied magnetic field.

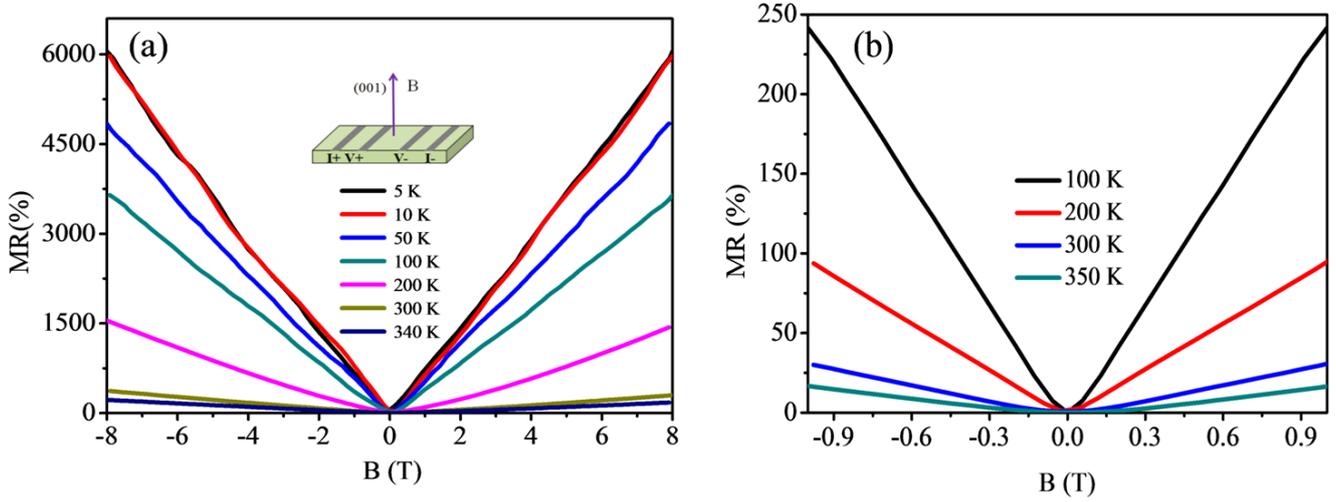

**Figure 2** (a) Magnetic field dependence of the MR in Bi$_{0.96}$Sb$_{0.04}$ single crystals at different temperatures. (b) Low-field linear MR in inhomogeneous Bi$_{0.96}$Sb$_{0.04}$ single crystals at different temperatures ranging from 100 K to 350 K.

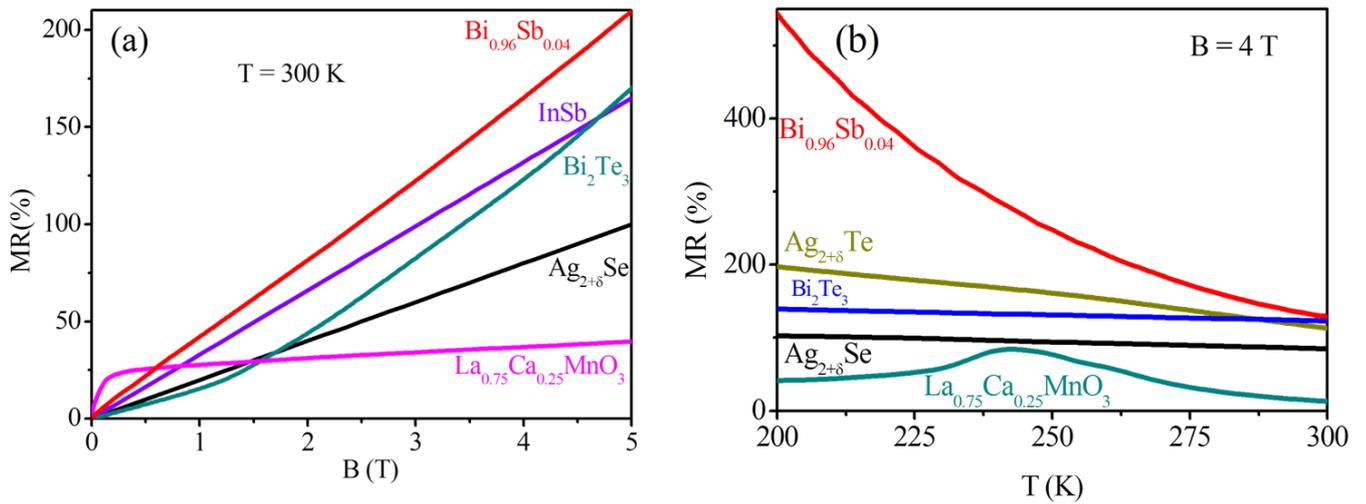

**Figure 3** (a) Comparison of magnetic field dependences of MR in Bi$_{0.96}$Sb$_{0.04}$ with those of Ag$_{2+\delta}$Se, Bi$_2$Te$_3$, InSb and La$_{0.75}$Ca$_{0.25}$MnO$_3$. (b) Comparison of temperature dependences of MR in Bi$_{0.96}$Sb$_{0.04}$ with those of Ag$_{2+\delta}$Se, Ag$_{2+\delta}$Te, Bi$_2$Te$_3$, and La$_{0.75}$Ca$_{0.25}$MnO$_3$.